\documentclass[twocolumn,superscriptaddres,prl]{revtex4}

\usepackage{graphicx}
\usepackage{epsfig}
\usepackage{amsmath, amssymb}
\usepackage{verbatim}

\begin{document}

\title{Anyon Condensation and Continuous Topological Phase Transitions \\
in Non-Abelian Fractional Quantum Hall States}

\author{Maissam Barkeshli}
\author{Xiao-Gang Wen}
\affiliation{Department of Physics, Massachusetts Institute of Technology,
Cambridge, MA 02139, USA }

\begin{abstract}

We find a series of possible continuous quantum phase transitions 
between fractional quantum Hall (FQH) states at the 
same filling fraction in two-component quantum Hall systems. 
These can be driven by tuning the interlayer tunneling and/or interlayer 
repulsion. One side of the transition is the Halperin $(p,p,p-3)$ Abelian
two-component state while the other side is the non-Abelian
$Z_4$ parafermion (Read-Rezayi) state. We predict that the
transition is a continuous transition in the 3D Ising class.
The critical point is described by a $Z_2$ gauged
Ginzburg-Landau theory. These results have implications for
experiments on two-component systems at $\nu = 2/3$ and
single-component systems at $\nu = 8/3$. 

\end{abstract}
%\pacs{PACS numbers: 73.43.-f, 11.25.Hf, 11.15.-q}

\maketitle

%\centerline{\Large Anyon Condensation and Topological Phase Transitions
%in Non-Abelian States}
%$\ $
%\centerline{Maissam Barkeshli and Xiao-Gang Wen}
%$\ $
%\centerline{Department of Physics, Massachusetts Institute of Technology,
%Cambridge, MA 02139, USA }
%$\ $
%\centerline{Nov. 7, 2009}

%$\ $

One of the most challenging problems in the study 
of quantum many-body systems is to understand transitions
between topologically ordered states \cite{Wen04}. Since topological
states cannot be characterized by broken symmetry and local order
parameters, we cannot use the conventional Ginzburg-Landau theory. 
When non-Abelian topological states are involved, the transitions that 
are currently understood are essentially all equivalent to the transition 
between weak and strong-paired BCS states \cite{W0050, RG0067}. Over the last ten years, while 
there has been much work on the subject, there has not been another quantum 
phase transition in a physically realizable system, involving a non-Abelian phase, 
for which we can answer the most basic questions of whether the transition can be continuous and
what the critical theory is. Here we present an additional example in the context of fractional
quantum Hall (FQH) systems.

The quasiparticle excitations in FQH states carry fractional statistics and fractional charge \cite{Wen04}.
%In bilayer fractional quantum Hall states, it is possible
%to tune the energy gap of these ``anyons'' to zero
%and induce a phase transition by tuning the interlayer 
%repulsion. 
In particular, in a $(ppq)$ bilayer FQH state \cite{Wen04, PG87},
there is a type of excitation, called a fractional exciton (f-exciton), which 
is a bound state of a quasiparticle in one layer and an oppositely charged 
quasihole in the other layer. It carries fractional statistics.
As we increase the repulsion between the electrons in
the two layers, the energy gap of the f-exciton will be reduced;
when it is reduced to zero, the f-exciton will condense and drive
a phase transition. When the anyon number has only a mod $n$ conservation, this can even lead 
to a non-Abelian FQH state \cite{W0050}, yet little is known
about ``anyon condensation'' \cite{BS0916,GA0901,BS0903,GS0903}. A better understanding of these phase transitions may
aid the quest for experimental detection of non-Abelian FQH states, because
one side of the transition -- in our case the 
$(330)$ state at $\nu = 2/3$ -- can be accessed experimentally \cite{SM9405,LJS9792}.
The results of this paper suggest a new way of experimentally tuning
to a non-Abelian state in bilayer FQH states, similar to the transition
from the $(331)$ state to the Moore-Read Pfaffian at $\nu = 1/2$ \cite{W0050,RG0067,PS1004}.

In the $(ppq)$ state, when the energy gap of the f-exciton 
at $\v k=0$ is reduced to zero, the f-exciton will condense\cite{W0050}. 
The transition can be described by the $\phi=0 \to \phi\neq 0$ transition in a Ginzburg-Landau 
theory with a Chern-Simons (CS) term: 
%\begin{align}
%\label{sCS} 
$\cL =  |(\prt_0 + i a_0 )\phi|^2 -  v^2 |(\prt_i + i a_i
)\phi|^2 - f |\phi|^2 - g |\phi|^4    -\frac{\pi}{\th}
\frac{1}{4\pi} a_\mu\prt_\nu a_\la \eps^{\mu\nu\la}, $
%\end{align}
where $\th$ is the statistical angle of the f-exciton.  Such a transition
changes the Abelian $(ppq)$ FQH state to another Abelian charge-$2e$
FQH state.\cite{W0050,RG0067}

In the presence of interlayer electron tunneling, the number
of f-excitons is conserved only mod $p-q$.  A new term $\del
\cL= t (\phi \hat M)^{p-q}+h.c.$ must be 
included, where $\hat M$ is an operator that creates $2\pi$
flux of the $U(1)$ gauge field $a_\mu$.  With this new term,
what is the fate of  the $\phi=0 \to \phi\neq 0$ transition?

When $p-q=2$, the f-excitons happen to be fermions (\ie
$\th=\pi$), so we can map the $\cL+\del\cL$
theory to a free fermion theory and solve the
problem.\cite{W0050} The problem is closely
related to the transition from weak to strong-pairing of a
$p_x + ip_y$ paired BCS superconductor \cite{RG0067}. 
The interlayer electron tunneling splits the single
continuous transition between the $(p,p,p-2)$ and the
charge-$2e$ FQH states into two continuous transitions.  The
new phase between the two new transitions is the non-Abelian
Pfaffian state \cite{MR9162}. This is the only class of phase
transition involving a non-Abelian FQH state for which
anything is known.

When $p-q\neq 2$, the f-excitons are anyons. The problem
becomes so hard that we do not even know where to start. But
we may guess that even when $p-q\neq 2$, an interlayer
electron tunneling may still split the transition between
the $(ppq)$ and charge-$2e$ FQH states. The new phase
between the two new transitions may be a non-Abelian FQH
state \cite{W0050}.  When $p-q=3$, it was
suggested that the new phase is a $Z_4$ parafermion
(Read-Rezayi\cite{RR9984}) FQH state \cite{RWR}.
This is because anti-symmetrizing the $(330)$ wave function between
the coordinates of the two layers yields the $Z_4$ parafermion
wave function, in direct analogy to the known continuous transition from $(331)$
to Pfaffian, where anti-symmetrizing the $(331)$ wave function
yields the single-layer Pfaffian wave function. 
%The reason is as follows.
%In the case of the $(331)$ to Pfaffian transition, 
%anti-symmetrizing the $(331)$ wave function between coordinates in the two layers
%yields the single-layer Moore-Read Pfaffian wave function. 
%This correctly suggests, at the level of wave functions, that increasing 
%the interlayer tunneling to infinity may yield a transition between
%the $(331)$ state to the Pfaffian state. Similarly, in \cite{RWR}, it was recently 
%observed that anti-symmetrizing the $(330)$ wave function between coordinates in the
%two layers yields the $Z_4$ parafermion wave function. 
%This raises some questions: can we have a continuous transition
%between the Abelian $(p,p,p-3)$ and the non-Abelian $Z_4$
%parafermion FQH states? What is the nature of the
%transition? What is the effective theory at the transition
%point?

In this letter, we show that \emph{the Abelian $(p,p,p-3)$
state can change into the $Z_4$ parafermion state through a
continuous quantum phase transition. The transition is in
the 3D Ising class.  The critical point is described by a
$Z_2$ gauged Ginzburg-Landau theory.} 
%The results here may
%help and guide us towards understanding an even broader
%class of transitions for which anyon condensation appears to
%play an important role, both for $p - q > 3$ and for more
%general multi-layer non-Abelian topological phase
%transitions driven by interlayer tunneling/repulsion.

These results are experimentally relevant in the case 
$p = 3$. The $(330)$ state has been experimentally realized 
in double layer and wide quantum wells \cite{LJS9792}. The existence 
of a neighboring single layer non-Abelian state in the 
phase diagram, which can be realized by tuning the interlayer 
tunneling/repulsion, suggests that experiments have a chance 
of realizing this transition. Furthermore, recent
detailed experimental studies of the energy gaps of the
$\nu = 8/3$ FQH state in single-layer systems indicate that it might be an exotic
state, as opposed to a conventional Laughlin or hierarchy state \cite{CK0801}.
This observation, together with the result of this paper that the $Z_4$ parafermion
state lies close to the experimentally observed $(330)$ state
in the quantum Hall phase diagram, suggests that the $Z_4$ parafermion
state ought to be considered as a candidate -- in addition to 
other proposed possibilities (e.g. \cite{BS0823}) -- in explaining the 
plateau at $\nu = 8/3$. In the case of the $5/2$ plateau,
numerical studies have recently suggested that finite layer thickness
of the quantum well may stabilize the non-Abelian Pfaffian state \cite{PJ0807}, while
it is also known that the Pfaffian state is near to the $(331)$ state
in the phase diagram; similarly, the results here suggest that finite
layer thickness may also help stabilize the $Z_4$ parafermion state at $\nu = 8/3$
because of its proximity in the phase diagram to the $(330)$ state. 

The conceptual breakthrough in our understanding is a recently 
discovered low energy effective theory for the $Z_4$ parafermion 
state, which was found to be a $U(1) \times U(1) \rtimes Z_2$ Chern-Simons
(CS) theory (a $U(1) \times U(1)$ CS theory coupled with a
$Z_2$ gauge symmetry).\cite{BW1023} This is closely related
to the effective theory for the $(p,p,p-3)$ state, which is
a $U(1) \times U(1)$ CS theory.  So effective
theories for the $Z_4$ and the $(p,p,p-3)$ states only
differ by a $Z_2$ gauge symmetry. Thus the transition
between the $(p,p,p-3)$ and the $Z_4$ states may just be a
$Z_2$ ``gauge symmetry-breaking'' transition induced by the
condensation of a $Z_2$ charged field.  We find that
the $U(1) \times U(1) \rtimes Z_2$ CS theory contains a
certain electrically neutral, bosonic quasiparticle that
carries a $Z_2$ gauge charge.  We argue that this bosonic
quasiparticle becomes gapless at the transition and its
condensation breaks the $Z_2$ gauge symmetry and yields the
$(p,p,p-3)$ state.

To obtain the above results, without losing generality, let
us choose $p=3$, and consider the $(330)$ state and the
corresponding filling fraction $\nu=2/3$ $Z_4$ parafermion
state. The same results would also apply to
filling fractions $\nu = 2n + 2/3$, where $n$ is an integer. 
We begin by explaining the quasiparticle content of
the $Z_4$ states, then we show that there exists an electrically
neutral bosonic quasiparticle in the $Z_4$ state whose
condensation yields the $(330)$ state and that carries a 
$Z_2$ gauge charge in the low energy effective theory.
Finally we discuss some consequences for physically measurable quantities.

One way to understand the topologically inequivalent excitations is through 
ideal wave functions, which admit a great variety of powerful 
tools for analysis of their physical properties \cite{MR9162,RR9984,WW0808,WW0809,BW0932,BW0937,
BH0802, BK0601,SL0604}.
%Another powerful approach is through the projective construction
%and the recently developed $U(1) \times U(1) \rtimes Z_2$ CS
%effective theory for the $Z_4$ state.\cite{Wnab,W9927,BW1002}
In the ideal wave function approach, the ground state and
quasiparticle wave functions of a FQH state are taken to be
correlation functions of a 2D CFT: $\Phi_\gamma(\{z_i\})
\sim \langle V_{\gamma}(0) \prod_{i=1}^{N} V_e(z_i)
\rangle$, where $\Phi_{\gamma}$ is a wave function with a
single quasiparticle of type $\gamma$ located at the origin
and $z_i = x_i + i y_i$ is the coordinate of the $i$th
electron.  $V_{\gamma}$ is a quasiparticle operator in the
CFT and $V_e$ are electron operators. The electron operator,
through its operator product expansions (OPE), forms the
chiral algebra of the CFT. Quasiparticles correspond to 
representations of the chiral algebra.  Two operators 
$V_{\gamma}$ and $V_{\gamma'}$ correspond to topologically 
equivalent quasiparticles if they differ by electron operators.

The $Z_4$ parafermion states, which exist at $\nu = 2/(2M+1)$ have
$5(2M+1)$ topologically distinct quasiparticles. These can be organized
into three representations of a magnetic translation
algebra,\cite{BW0932} which each contain $2(2M+1)$,
$2(2M+1)$, and $2M+1$ quasiparticles -- see Table \ref{Z4paraQP}, 
where we also listed a representative operator in the corresponding
CFT description of these states. The CFT description of these states
is formulated in terms of the $Z_4$ parafermion CFT \cite{ZF8515} and a free
boson CFT. The $Z_4$ parafermion CFT can be formulated in terms of
an $SU(2)_4/U(1)$ coset CFT \cite{GQ8723} or, equivalently, as
the theory of a scalar boson $\varphi_r$, compactified at a special 
radius $R^2 = 6$ so that $\varphi_r \sim \varphi_r + 2\pi R$,
and that is gauged by a $Z_2$ action: $\varphi_r \sim - \varphi_r$ \cite{DV8985}. Such a
CFT is called the $U(1)/Z_2$ orbifold CFT. In Table \ref{Z4paraQP}, we have included
labellings of the operators in the CFT using both of these formulations. 

\begin{comment}
\begin{table}
\begin{tabular}{clc}
\multicolumn{3}{c}{Quasiparticles for $Z_4$ state at $\nu = 2/3$} \\
\hline
& CFT Label & Scaling Dimensions \\
\hline
$\v 0$ & $\mathbb{I} \sim \phi_N^1 e^{i\sqrt{3/2}\varphi_c}$ & 0+0 \\
$\v 1$ & $e^{i 2/3 \sqrt{3/2} \varphi_c}$ & 0+1/3 \\
$\v 2$ & $e^{i 4/3 \sqrt{3/2} \varphi_c}$ & 0+4/3 \\
$\v 3$ & $j \sim \partial \varphi_r$ & 0+1 \\
$\v 4$ & $j e^{i 2/3 \sqrt{3/2} \varphi_c}$ & 0+4/3 \\
$\v 5$ & $j e^{i 4/3 \sqrt{3/2} \varphi_c}$ & 0+7/3 \\
\\
$\v 6$ & $\sigma_1 e^{i 1/6 \sqrt{3/2} \varphi_c}$ & 1/16 + 1/48 \\
$\v 7$ & $\sigma_1 e^{i 5/6 \sqrt{3/2} \varphi_c}$ & 1/16 + 25/48 \\
$\v 8$ & $\sigma_1 e^{i 9/6 \sqrt{3/2} \varphi_c}$ & 1/16 + 27/16 \\
$\v 9$ & $\tau_1 e^{i 1/6 \sqrt{3/2} \varphi_c}$ & 9/16 + 1/48 \\
$\v{10}$ & $\tau_1 e^{i 5/6 \sqrt{3/2} \varphi_c}$ & 9/16 + 25/48 \\
$\v{11}$ & $\tau_1 e^{i 9/6 \sqrt{3/2} \varphi_c}$ & 9/16 + 27/48 \\
\\
$\v{12}$ & $\cos(\frac{\varphi}{\sqrt{6}}) e^{i 1/3 \sqrt{3/2} \varphi_c}$ & 1/12 + 1/12 \\
$\v{13}$ & $\cos(\frac{\varphi}{\sqrt{6}}) e^{i \sqrt{3/2} \varphi_c}$  & 1/12 + 3/4 \\
$\v{14}$ & $\cos(\frac{\varphi}{\sqrt{6}}) e^{i 5/3 \sqrt{3/2} \varphi_c}$  & 1/12 + 25/12 \\
\hline
\end{tabular}
\end{table}
\end{comment}

\begin{table}
\begin{tabular}{cllcc}
%\hline
%\multicolumn{5}{|c|}{Quasiparticles for $Z_4$ state with $M = 1$ ($\nu = 2/3$)} \\
\hline
 & $\Phi^l_m e^{i Q \sqrt{\nu^{-1}} \varphi_c}$ & $Z_2$ Orbifold Label & $\{n_l\}$ & $h_{pf} + h_{ga}$ \\
\hline
$\v 0$ & $\mathbb{I} \sim \Phi^0_2 e^{i \sqrt{3/2} \varphi_c} $ & $\mathbb{I} \sim \phi_N^1 e^{i \sqrt{3/2} \varphi_c} $ & 1 1 1 0 0 1 & 0+0 \\
%\hline
$\v 1$ & $e^{i 2/3 \sqrt{3/2} \varphi_c}$ & $e^{i 2/3 \sqrt{3/2} \varphi_c}$ & 1 1 1 1 0 0 & $0+\frac{1}{3}$ \\
%\hline
$\v 2$ & $e^{i 4/3 \sqrt{3/2} \varphi_c}$ & $e^{i 4/3 \sqrt{3/2} \varphi_c}$ & 0 1 1 1 1 0 & $0+\frac{4}{3}$ \\
%\hline
$\v 3$ & $\Phi^0_4$ & $j_r \sim \partial \varphi_r$ & 0 0 1 1 1 1 & 1+0 \\
%\hline
$\v 4$ & $\Phi^0_4 e^{i 2/3 \sqrt{3/2} \varphi_c}$ & $j_r e^{i 2/3 \sqrt{3/2} \varphi_c}$ & 1 0 0 1 1 1 & $1+\frac{1}{3}$ \\
$\v 5$ & $\Phi^0_4 e^{i 4/3 \sqrt{3/2} \varphi_c}$ & $j_r e^{i 4/3 \sqrt{3/2} \varphi_c}$ & 1 1 0 0 1 1 & $1+\frac{4}{3}$ \\
%& & & & \\
\hline
$\v 6$ & $\Phi^1_1 e^{i 1/6 \sqrt{3/2} \varphi_c}$ & $\sigma_1 e^{i 1/6 \sqrt{3/2} \varphi_c}$ & 1 1 0 1 0 1 & $\frac{1}{16} + \frac{1}{48}$ \\
$\v 7$ & $\Phi^1_1 e^{i 5/6 \sqrt{3/2} \varphi_c}$ & $\sigma_1 e^{i 5/6 \sqrt{3/2} \varphi_c}$ & 1 1 1 0 1 0 & $\frac{1}{16} + \frac{25}{48}$ \\
$\v 8$ & $\Phi^1_{1} e^{i 9/6 \sqrt{3/2} \varphi_c}$ & $\sigma_1 e^{i 9/6 \sqrt{3/2} \varphi_c}$ & 0 1 1 1 0 1 & $\frac{1}{16} + \frac{27}{16}$ \\
$\v 9$ & $\Phi^1_5 e^{i 1/6 \sqrt{3/2} \varphi_c}$ & $\tau_1 e^{i 1/6 \sqrt{3/2} \varphi_c}$ & 1 0 1 1 1 0 & $\frac{9}{16} + \frac{1}{48}$ \\
$\v{10}$ & $\Phi^1_5 e^{i 5/6 \sqrt{3/2} \varphi_c}$ & $\tau_1 e^{i 5/6 \sqrt{3/2} \varphi_c}$ & 0 1 0 1 1 1 & $\frac{9}{16} + \frac{25}{48}$ \\
$\v{11}$ & $\Phi^1_5 e^{i 9/6 \sqrt{3/2} \varphi_c}$ & $\tau_1 e^{i 9/6 \sqrt{3/2} \varphi_c}$ & 1 0 1 0 1 1 & $\frac{9}{16} + \frac{27}{16}$ \\
%& & & & \\
\hline
$\v{12}$ & $\Phi^2_2 e^{i 1/3 \sqrt{3/2} \varphi_c}$ & $\cos(\frac{\varphi_r}{\sqrt{6}}) e^{i 1/3 \sqrt{3/2} \varphi_c}$  & 1 0 1 1 0 1 & $\frac{1}{12} + \frac{1}{12}$ \\
$\v{13}$ & $\Phi^2_2 e^{i \sqrt{3/2} \varphi_c}$ & $\cos(\frac{\varphi_r}{\sqrt{6}}) e^{i \sqrt{3/2} \varphi_c}$ & 1 1 0 1 1 0 & $\frac{1}{12} + \frac{3}{4}$ \\
$\v{14}$ & $\Phi^2_2 e^{i 5/3 \sqrt{3/2} \varphi_c}$ & $\cos(\frac{\varphi_r}{\sqrt{6}}) e^{i 5/3 \sqrt{3/2} \varphi_c}$ & 0 1 1 0 1 1 & $\frac{1}{12} + \frac{25}{12}$\\
\hline
%\multicolumn{5}{c}{} \\
%\hline
%\multicolumn{5}{|c|}{Quasiparticles for $Z_4$ state with $M = 0$ ($\nu = 2$)} \\
%\hline
%0 & $\mathbb{I} \sim \Phi^0_2 e^{i\sqrt{2}\varphi_c}$ & $\mathbb{I} \sim \phi_N^1 e^{i \sqrt{3/2} \varphi_c} $ & 4 0 & 0 \\
%1 & $\Phi^0_4$ & $j_r$ & 0 4 & 1 \\
%& & & & \\
%2 & $\Phi^1_1 e^{i (1/2) \sqrt{1/2}\varphi_c}$ & $\sigma_1 e^{i 1//2 \sqrt{2} \varphi_c}$ & 3 1 & $\frac{1}{16} + \frac{1}{16}$ \\
%3 & $\Phi^1_5 e^{i (1/2) \sqrt{1/2}\varphi_c}$ & $\tau_1 e^{i 1/2 \sqrt{2} \varphi_c}$ & 1 3 & $\frac{9}{16} + \frac{1}{16}$ \\
%& & & & \\
%4 & $\Phi^2_2 e^{i \sqrt{1/2} \varphi_c}$ & $\cos(\frac{\varphi_r}{\sqrt{6}}) e^{i \sqrt{2} \varphi_c}$ & 2 2 & $\frac{1}{12} + \frac{1}{4}$ \\
%\hline
\end{tabular}
\caption{\label{Z4paraQP}
Quasiparticles in the $Z_4$ parafermion FQH state at $\nu =
2/3$ ($M = 1$). The different representations of the
magnetic translation algebra\cite{BW0932} are separated by
horizontal lines. $Q$ is the electric charge and $h_{pf}$
and $h_{ga}$ are the scaling dimensions of the $Z_4$
parafermion field $\Phi^l_m$ and the $U(1)$ vertex operator
$e^{i \alpha \varphi_c}$, respectively. $\varphi_c$ is a
free scalar boson that describes the charge sector.
$\{n_l\}$ is the occupation number sequence associated with
the quasiparticle pattern of zeros. }
\end{table}

The fusion rules of these quasiparticles can be obtained
from the fusion rules of the $Z_4$ parafermion CFT:
$\Phi^0_a \times \Phi^l_m = \Phi^l_{m+a}$, $\Phi^1_1 \times \Phi^1_1
= \Phi^0_2 + \Phi^2_2$, and $\Phi^2_2 \times \Phi^2_2 =
\mathbb{I} + \Phi^0_4 + \Phi^2_0$.  $\Phi^l_m$ exists for
$l+m$ even, $0 \leq l \leq n$, and is subject to the
following equivalences: $\Phi^l_m \sim \Phi^l_{m+2n} \sim
\Phi^{n-l}_{m-n}$, where $n = 4$ for the $Z_4$ parafermion
CFT.

%$n = 4$ for the $Z_4$ parafermion CFT. The scaling dimension, $h^l_m$, of the operator
%$\Phi^l_m$ is given by:
%\begin{align}
%\label{parafermionWeight}
%h^l_m = \left\{
%  \begin{array}{lll}
%    \frac{l(l+2)}{4(n+2)} - \frac{m^2}{4n} + \frac{m-l}{2} &\mbox{ if } & l \leq m \leq 2n - l \\
%    \frac{l(l+2)}{4(n+2)} - \frac{m^2}{4n} & \mbox{ if } &-l \leq m \leq l \\
%    \end{array} \right.
%\end{align}

Another useful way to understand the topological order of
the $Z_4$ FQH state is through its bulk
effective field theory, for which there are several
different formulations.\cite{CF0057,BW1023,BW1002} Here
we use the $U(1) \times U(1) \rtimes
Z_2$ CS theory.\cite{BW1023} This is the theory of two
$U(1)$ gauge fields, $a$ and $\tilde{a}$, with an additional
$Z_2$ gauge symmetry that corresponds to interchanging $a$
and $\tilde{a}$ at any given point in space. The Lagrangian is
$\mathcal{L} = \frac{p}{4\pi} (a\partial a + \tilde{a} \partial \tilde{a}) +
\frac{q}{4\pi} (a \partial \tilde{a} + \tilde{a} \partial a)$,
where $a \partial a$ is shorthand for $\epsilon^{\mu \nu
\lambda} a_\mu \partial_\nu a_\lambda$.  For $p-q = 3$, it
was found that the ground state degeneracy on genus $g$
surfaces of this theory agrees with that of the $Z_4$
parafermion FQH states at $\nu = 2/(2p-3)$.  In addition, it
was found that the $Z_2$ vortices, which correspond to
defects around which $a$ and $\tilde{a}$ are interchanged,
correspond to the non-Abelian $\Phi^1_m$ quasiparticles ($\v
6 - \v{11}$ in Table \ref{Z4paraQP}).\cite{BW1023} It was
also found that quasiparticle $\v 3$, which corresponds to
the operator $\Phi^0_4$ (see Table \ref{Z4paraQP}), is
charged under the $Z_2$ gauge symmetry.  This latter result is
suggested by the orbifold formulation of the $Z_4$
parafermion CFT \cite{DV8985}, where $\Phi^0_4$ corresponds to a $U(1)$
current $j_r \sim \partial \varphi_r$.  In the $Z_2$
orbifold, the scalar boson $\varphi_r$ is gauged by the
$Z_2$ action $\varphi_r \sim - \varphi_r$. The $Z_2$ gauge
symmetry in the bulk CS theory is the $Z_2$ gauging in the orbifold
CFT, which suggests that the quasiparticle
$\Phi^0_4$ would carry a $Z_2$ gauge charge.

From Table \ref{Z4paraQP}, we see that the $Z_4$ states
contain a special quasiparticle, $\v 3$ ($\Phi^0_4$), which
is electrically neutral, fuses with itself to the identity,
and has Abelian fusion rules with all other quasiparticles.
$\Phi^0_4$ has scaling dimension 1 and is a bosonic operator.
%Thus it is local with respect to itself; the corresponding operator in
%the CFT, $\Phi^0_4$, has scaling dimension 1 and satisfies the OPE:
%$\Phi^0_4(z) \Phi^0_4(w) \sim \frac{1}{(z-w)^2} + \cdots$.
In the following we show that the condensation of this
neutral bosonic quasiparticle yields the topological order
of the $(330)$ phase.  Before condensation, two excitations
are topologically equivalent if they differ by an electron,
which is a local excitation. After condensation, all allowed
quasiparticles must be local with respect to both the
electron and $\Phi^0_4$, and two quasiparticles will be
topologically equivalent if they differ either by an
electron or by $\Phi^0_4$. In the CFT language, this means
that $\Phi^0_4$ has been added to the chiral algebra and will
appear in the Hamiltonian.  Such a situation was analyzed in
a general mathematical setting for topological phases in \Ref{BS0916}.

From the OPE of $\Phi^0_4$ and the other quasiparticle
operators in the CFT description, we find that $\Phi^0_4$ is
mutually local with respect to the quasiparticles in the
first and third representations of the magnetic algebra,
which consist of the quasiparticles made of $\Phi^0_m$ and
$\Phi^2_m$ (see Table \ref{Z4paraQP}). However, its mutual
locality exponent with the $\Phi^1_m$ quasiparticles is
half-integer, which means that $\Phi^0_4$ is non-local with
respect to those quasiparticles. This is expected, because
the $\Phi^1_m$ quasiparticles were found to correspond to
$Z_2$ vortices in the $U(1) \times U(1) \rtimes Z_2$ CS
theory while $\Phi^0_4$ was found to carry $Z_2$ charge.
Thus we would expect that $\Phi^0_4$ would be non-local
with respect to the $\Phi^1_m$ quasiparticles, with a
half-integer mutual locality exponent. As a result,
quasiparticles $\v 6 - \v{11}$ are no longer valid
(particle-like) topological excitations after condensation.

Since quasiparticles that differ by $\Phi^0_4$ are regarded
as topologically equivalent after condensation,
quasiparticles $\v 0$, $\v 1$, and $\v 2$ become
topologically equivalent to quasiparticles $\v 3$, $\v 4$,
and $\v5$ (see Table \ref{Z4paraQP}), leaving three
topologically distinct quasiparticles from this
representation.
%This leaves a total of $6$ quasiparticles, which correspond
%to the 6 quasiparticles of the $(330)$ state that are invariant under a global $Z_2$ symmetry.
Furthermore, the three quasiparticles in the third
representation split into 6 topologically distinct
quasiparticles. The reason for this was discussed in
\Ref{BS0916}. Consider the fusion of a $\Phi^2_2$
quasiparticle, which we will label as $\v{\gamma}$, and its
conjugate: $\v{\gamma} \times \v{\bar{\gamma}} = \v 0 + \v 3
+ \v{13}$.  After condensation, we identify $\v 3$
($\Phi^0_4$) with the vacuum sector, so if $\v \gamma$ does
not split into at least two different quasiparticles, then
there would be two different ways for it to annihilate into
the vacuum with its conjugate. A basic property of
topological phases is that particles annihilate into the
vacuum in a unique way, so $\v \gamma$ must split into at
least two different quasiparticles. Since the quantum
dimension of $\v \gamma$ is $2$, it must split into exactly
two quasiparticles, each with quantum dimension 1: $\v
\gamma \rightarrow \v \gamma_1 + \v \gamma_2$. $\v \gamma_1$
and $\v \gamma_2$ are now Abelian quasiparticles because
they have unit quantum dimension. Therefore, we see that the
15 quasiparticles in the $Z_4$ parafermion state become,
after condensation of $\Phi^0_4$, the 9 Abelian
quasiparticles of the $(330)$ state.

As the energy gap to quasiparticle 
$\v 3$ ($\Phi^0_4$) is reduced, the low energy effective theory will
simply be the theory of this bosonic field coupled to a
$Z_2$ gauge field. The phase transition is a 
Higgs transition of this $Z_2$ charged boson (at least in the case 
where we explicitly break the global $Z_2$ symmetry of layer exchange). 
Note that there is no $U(1)$ symmetry that conserves the density 
of this kind of excitation because $\Phi^0_4$ can annihilate 
with itself into the vacuum.  Such a theory of a real 
scalar coupled to a $Z_2$ gauge field was studied
in \Ref{FS7982}, and it was found that the transition is
continuous and in the 3D Ising universality class.  As we
mentioned before, the same transition, when viewed as a
transition from the $(330)$ state to the $Z_4$ state, is
induced by an anyon condensation with a mod-3 conservation.
This suggests that the anyon condensation can, surprisingly, 
be described by a $Z_2$-charged boson condensation.

Given this result, a natural question is why fluctuations
of $\Phi^0_4$ should physically be related to interlayer density 
fluctuations, which can be tuned by the interlayer tunneling
and interlayer repulsion. One answer to this question comes
from the analysis of the wave functions. The $(330)$ wave function
in real space is 
$\Phi_{(330)}(\{z_i,w_i\}) = \langle 0 | \prod_{i=1}^N \psi_{e1}(z_i) \psi_{e2}(w_i) | 330\rangle$,
where $\psi_{ei}$ is the electron annihilation operator in the $i$th layer. 
When interlayer tunneling is increased, the gap between the single-particle symmetric
and anti-symmetric states is increased, until eventually all 
electrons occupy the symmetric set of orbitals, created by $\psi_{e+}$, where
$\psi_{e\pm} \propto \psi_{e1}^\dagger \pm \psi_{e2}^\dagger$. The natural wave function
to guess in the limit of infinite interlayer tunneling is thus
the projection onto the symmetric states:
$\Phi(\{z_i\}) = \langle 0 | \prod_{i=1}^{2N} (\psi_{e1}(z_i) + \psi_{e2}(z_i)) |330\rangle$,
which in this case is the $Z_4$ parafermion wave function \cite{RWR}.
Thus starting from the $(330)$ state, the wave function analysis suggests that interlayer tunneling 
will yield the $Z_4$ parafermion state. Since the $(330)$ state
can be obtained from the $Z_4$ parafermion state by condensing
$\Phi^0_4$, we are led to take this wave function analysis seriously and are
led to conclude that the gap to $\Phi^0_4$ can be controlled by interlayer tunneling. 
Note that tuning interlayer tunneling will tune fluctuations of the operator
$\psi_{e+}^\dagger \psi_{e-}$, which is related to interlayer
density fluctuations (since $\psi_{e+}^\dagger \psi_{e-} + h.c. = \psi_{e1}^\dagger \psi_{e1} - \psi_{e2}^\dagger \psi_{e2}$).
Since the dimensionless parameters that we are concerned with
are $t/V_{inter}$ and $V_{inter}/V_{intra}$, we see that properly tuning
the inter/intra -layer Coulomb repulsions (given by $V_{inter/intra}$) 
should also drive the interlayer density fluctuations and the corresponding transition. 
For yet a different perspective, we refer to\cite{BW101}. 

Since the transition between these two FQH states is driven
by the condensation of a neutral quasiparticle, it will be
difficult to observe in experiments.  Experiments
have observed a phase transition in bilayer systems at $\nu
= 2/3$ by measuring the kink in the charged excitation gap
in charge transport measurements, though
that may be an indication of a different transition \cite{LJS9792}.
One can also observe the $(330)$ to $Z_4$ parafermion transition directly by
probing the gapless neutral mode. The analysis here predicts
that the bulk of the sample should remain an electrical insulator but
become a thermal conductor at the transition. Furthermore, as the transition is
approached, the fluctuations in the operator
$\psi_{e-}^{\dagger} \psi_{e+}$ should correspond to
fluctuating electric dipole moments between the two layers,
which can be probed through surface acoustic
waves.\cite{WR9344} Alternatively, experiments measuring minimal
quasiparticle charge would be able to detect this transition,
because the $Z_4$ state at $\nu = 2/3$ has a minimal 
quasiparticle charge of $e/6$, while the $(330)$ state has 
a minimal quasiparticle charge of $e/3$.

This research is supported by NSF Grant No.  DMR-0706078.

%\bibliography{Z2trans,../../bib/all,../../bib/wencross,../../bib/publst}
%\bibliography{/Users/Maissam/Documents/Research/biblio/biblio,all,wencross,publst}

\end{document}